\documentstyle[twocolumn,aps]{revtex}
\begin{document}
\input psfig
\draft

%
\twocolumn[\hsize\textwidth\columnwidth\hsize\csname@twocolumnfalse\endcsname
%
%

\title{Magnetic Domains and Stripes in the Spin-Fermion Model for Cuprates}

\author{Charles Buhler, Seiji Yunoki and Adriana Moreo}

\address{Department of Physics, National High Magnetic Field Lab and
MARTECH,\\ Florida State University, Tallahassee, FL 32306, USA}

\date{\today}
\maketitle

\begin{abstract}

Monte Carlo simulations applied to the Spin-Fermion model for cuprates 
show the existence of antiferromagnetic spin domains and charge stripes
upon doping. The stripes are partially filled, with a filling of
approximately 1/2 hole per site,
and they separate spin domains with a $\pi$ phase shift among them.
The stripes observed run either along the x or y axes
and they are separated
by a large energy barrier. No special boundary conditions or external fields
are needed to stabilize these structures at low temperatures. 
When magnetic incommensurate 
peaks are observed at momentum $\pi(1,1-\delta)$ and symmetrical points,
charge incommensurate peaks appear at $(0,2 \delta)$ and symmetrical
points, as experimentally observed. The strong charge fluctuations
responsible for the formation of the stripes also induce a pseudogap in the
density of states.

\end{abstract}

\vskip2pc]
\narrowtext

In recent years neutron scattering experiments have established that magnetic
incommensurability is a property common to most of the high-Tc
cuprates\cite{Cheong}. In addition, there is mounting evidence
supporting charge stripe formation in these compounds 
as well.\cite{Tran} These nontrivial spin and charge arrangements may be
 crucial to understand the unusual transport and superconducting
behavior of the cuprates.
Early Hartree-Fock studies of the
Hubbard model already predicted stripe formation with
insulating characteristics.\cite{zaanen}
 Phase-separation between hole-rich and
hole-poor regions in the
$\rm CuO_2$ planes, supplemented by long-range Coulomb interactions,
has also been proposed to explain the existence of stripes.\cite{Eme}
In addition, ground states with metallic stripes made out of
d-wave hole pairs 
have been observed in the t-J model using special
boundary conditions.\cite{white} 
However, a more detailed
theoretical understanding of these phenomena and resolution of current
conflicting results have been extremely
challenging especially since the t-J and Hubbard models
used for the cuprates are
considerably difficult to study. 
As an alternative to this more traditional approach here we  present the first
computational study of a simpler phenomenological model, the
Spin-Fermion (SF) model, which has been previously analyzed mostly using mean
field approximations with the main
goal of understanding d-wave 
superconductivity.\cite{Pines,Schrieffer,Fedro}
Here the focus is instead shifted toward the magnetic and
charge properties of the SF model, which have not been explored before
using unbiased techniques. 
In carrying out such a study
unexpected results were observed, notably the presence of
spin incommensurability and
metallic stripe formation at finite hole density, in excellent agreement
with experiments.

The SF model is constructed as an interacting system of
electrons and spins, mimicking phenomenologically the
coexistence of charge and spin degrees of freedom in 
the cuprates.\cite{Pines,Schrieffer,Fedro}. Its Hamiltonian is given by

$$
{\rm H=
-t{ \sum_{\langle {\bf ij} \rangle\alpha}(c^{\dagger}_{{\bf i}\alpha}
c_{{\bf j}\alpha}+h.c.)}}
+{\rm J
\sum_{{\bf i}}
{\bf{s}}_{\bf i}\cdot{\bf{S}}_{\bf i}
+J'\sum_{\langle {\bf ij} \rangle}{\bf{S}}_{\bf i} \cdot{\bf{S}}_{\bf j}},
\eqno(1)
$$
\noindent where ${\rm c^{\dagger}_{{\bf i}\alpha} }$ creates an electron
at site ${\bf i}=(i_x,i_y)$ with spin projection $\alpha$,  
${\bf s_i}$=$\rm \sum_{\alpha\beta} 
c^{\dagger}_{{\bf i}\alpha}{\bf{\sigma}}_{\alpha\beta}c_{{\bf
i}\beta}$ is the spin of the mobile electron, the  Pauli
matrices are denoted by ${\bf{\sigma}}$,
${\bf{S}_i}$ is the localized
spin at site ${\bf i}$,
${ \langle {\bf ij} \rangle }$ denotes nearest-neighbor (NN)
lattice sites,
${\rm t}$ is the NN-hopping amplitude for the electrons,
${\rm J>0}$ is an antiferromagnetic (AF) coupling between the spins of
the mobile and localized degrees of freedom (DOF),
and ${\rm J'>0}$ is a direct AF coupling
between the localized spins.
The density $\rm \langle n \rangle$=$\rm 1-x$ of 
itinerant electrons is controlled by a chemical potential $\mu$. 
Hereafter ${\rm t=1}$ will be used as the unit of energy. From 
previous phenomenological analysis the coupling ${\rm J}$ 
is expected to be larger than t, while the Heisenberg coupling
${\rm J'}$ is expected to be smaller.\cite{Schrieffer,Fedro} 

Here classical spins with $|S_{\bf i}|=1$ will be used for the 
localized spins, as also assumed in previous literature.\cite{foot}
This will allow us to
perform Monte Carlo (MC) simulations of model Eq.(1)
without ``sign problems'', reaching by this procedure
temperatures as low as T=0.01 at
any density.
This temperature is well below T=0.2, the lowest that can be
stabilized with quantum MC away from half-filling in the standard 
Hubbard model.~\cite{adri}
The value of ${\rm J}$ will be fixed to 2,
as suggested in Ref.~\cite{Fedro} 
where comparisons with experimental results
were performed. 
The coupling ${\rm J'}$ among the
classical spins will be set to 0.05. This value was selected by
monitoring the magnetic susceptibility and comparing its behavior to
experimental results for the cuprates.
The present study has
been performed mostly on 8$\times$8 
lattices with periodic boundary conditions
(PBC), but occasional runs were made 
also using open and antiperiodic BC
as well as different lattice sizes.
The numerical technique used here involves a standard Metropolis
algorithm for the classical spins and an exact diagonalization for the
itinerant electrons. The details of the method have been 
described in Ref.~\cite{yuno}. 
 
To study the magnetic properties of the system we measured the spin-spin
correlation functions among the classical spins defined as
$\rm \omega({\bf r})={1\over{N}}\sum_{\bf i}\langle{\bf{S}}_{\bf
i}\cdot{\bf{S}}_{\bf i+r}\rangle,$
where N is the number of sites.
The Fourier transform of $\rm \omega({\bf r})$, i.e. the spin structure
factor $S({\bf q})$, was
also investigated.
The momentum $\rm q_\gamma$ takes the values ${\rm 2 \pi
n/L_\gamma}$, with
${\rm n}$ running from 0 to ${\rm L_\gamma-1}$, and ${\rm L_\gamma}$ denoting 
the number of sites
along the $\gamma$=x or y direction. 
In our MC simulations long-range AF
order has been observed at $\rm \langle n \rangle$=1.0 as expected. 
As the
electron density is reduced from 1, the $S(\pi,\pi)$ intensity decreases.
One of the main results observed in this effort occurs at finite hole
density where a remarkable spin $incommensurability$ 
appears at $\rm \langle n \rangle$$\approx$0.8 with the 
structure factor peak moving to
$\pi((1-\delta),1)=(3\pi/4,\pi)$ and 
rotated points.
This behavior
is illustrated in Fig.1-a where $S({\bf q})$ on  
8$\times$8 lattices at T=0.01
is shown for different
values of $\rm \langle n \rangle$ along
the path $(0,0)-(0,\pi)-(\pi,\pi)-(0,0)$. Results along 
$(0,0)-(\pi,0)-(\pi,\pi)-(0,0)$ are in the inset. 
\begin{figure}[thbp]
\centerline{\psfig{figure=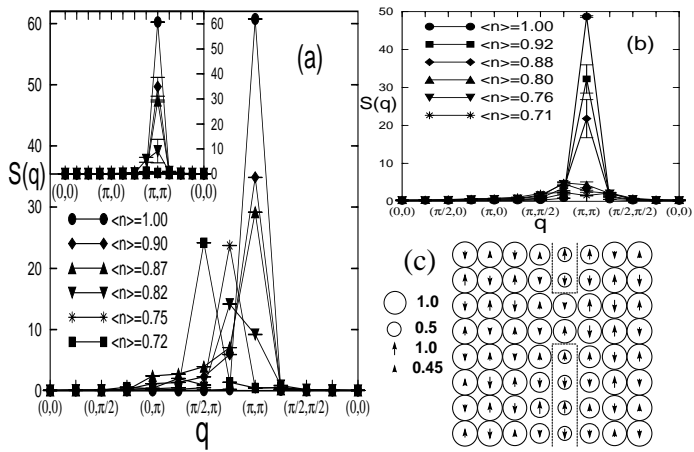,height=6.5cm}}
\caption{(a) Structure factor $S({\bf q})$ for the localized spins versus
momentum for J=2, $\rm J'$=0.05 and T=0.01 on an 8$\times$8
lattice and several densities along
$(0,0)-(0,\pi)-(\pi,\pi)-(0,0)$ (inset: results along
$(0,0)-(\pi,0)-(\pi,\pi)-(0,0)$)); (b) same as (a) but at T=0.05; (c) spin and
charge distribution for a typical MC 
snapshot at $\rm \langle n \rangle$=0.9 and the
same parameters as in (a). The arrow lengths are proportional to the local spin
$S_z({\bf i})$ and
the radius of the circles is proportional to the local density
$n({\bf i})$, according to the scale
shown. PBC are used.}
\end{figure}
Note that the spin correlations
along the two paths of Fig.1-a are different because at low temperatures
the symmetry under lattice $\pi/2-$rotations appears spontaneously $broken$. 
Similar ground-state properties but rotated by $\pi/2$ have also been
observed in independent runs depending on the initial conditions\cite{foot1} 
for the classical spins indicating that there are two energy minima
separated by a large barrier.
At T=0.05 (Fig.1-b) and higher temperatures the rotational symmetry is
restored but the intensity of the
incommensurate (IC) peaks is considerably reduced
(e.g. at $\rm \langle n \rangle$=0.75 the T=0.01
IC peaks are four times higher than at T=0.05).
This may explain the low intensity of the IC peaks 
observed in the T=0.2 Hubbard model quantum MC simulations.\cite{adri}

Experimentally it was observed that $\delta\approx$0.25 corresponds to the
saturation value reached at hole density x$\approx$0.12, which persists up to
x$\approx$0.25.\cite{yamada}
 Due to the finite size of the
lattices studied here, incommensurability for values of $\delta$ smaller
than 0.25 cannot be detected apparently preventing us from comparing directly with the
low-doping
experimental results. However, we have analyzed
typical spin
configurations (snapshots) emerging from our MC simulations close to
$\rm \langle n \rangle$=1.0
and we observed the existence of large AF
spin domains in most of them, an example of which is shown in Fig.1-c
for $\rm \langle n \rangle$=0.9.
The existence of these domains clearly suggests that tendencies towards 
magnetic
incommensurability appear in the system at x$\leq$0.20 as well.


The origin of the short-range incommensurate magnetic order in the
cuprates is still not clear. Magnetic order due to charge order
has been proposed as a possible explanation. 
To explore the possibility of charge ordering in the SF model,
$N({\bf q})$ was studied here defined as 
the Fourier transform 
of the charge correlations 
$\rm  n({\bf r})={1\over{N}}\sum_{\bf i} \langle (n_{\bf i}-\langle n \rangle)(
n_{\bf i+r}-\langle
n \rangle)\rangle ,$
\noindent where $\rm n_{\bf i}$ is the number operator at site ${\bf i}$
for the itinerant fermions.
In the AF phase at
$\rm \langle n \rangle$=1.0 $N({\bf q})$ was observed to present a broad peak
at ${\bf q}=(\pi,\pi)$ due to negative charge 
correlations (charge repulsion) 
at very short distances,
in agreement with previous calculations~\cite{chino}. As the system is
hole doped the behavior of $N({\bf q})$ 
becomes more temperature dependent. At
T=0.05 the peak in $N({\bf q})$ remains at
$(\pi,\pi)$, but at T=0.01 a sharp peak appears at small 
momenta for 0.7$\rm < \langle n \rangle <$0.9
(Figs.2-a,b) indicating the existence of extended charge structures.
If charge and spin incommensurability were related, stripe 
studies\cite{Eme} predict that the peak in $N({\bf q})$ has to appear at 
$(2 \delta,0)$ and symmetrical points. Then, incommensurability
associated to $\delta$=0.125 that cannot be explicitly
 detected in $S({\bf q})$
due to the size of our clusters can nevertheless
 be observed in the charge channel.
Indeed in Fig.2-a the peak in $N({\bf q})$ for density 0.87 is located at
${\bf q}=(\pi/4,0)$ compatible with $~\delta\approx 0.125$. At low temperature
the peak can
be observed along the x direction but not along y (see inset of Fig.2-a)
indicating that in the MC runs described here
the charge domains are along the y direction, causing
the spontaneous breaking of rotational symmetry described before. 
This clearly can be seen in the
charge distribution MC snapshot at $\rm \langle n \rangle$$\approx$0.85 shown
in Fig.3-a. Note that the holes are located along the magnetic
domain boundaries. The presence of stripes was explicitly verified also
using 12$\times$12 clusters.
The result is in excellent agreement with
experiments\cite{Tran1} on Nd-doped LSCO where vertical stripes are
observed at the Sr concentration $\rm x\approx 1/8$. 

In some stripe scenarios\cite{Zachar} charge ordering is expected to occur
at higher temperature than magnetic order. However, in model Eq.(1) both charge
and magnetic ordering appear to occur at similar
temperatures. Consider, for example, in 
Fig.3-b a typical MC snapshot at $\rm \langle n \rangle$=0.75 and
T=0.01. The holes are here aligned along $two$ vertical columns.
$S({\bf q})$ and $N({\bf q})$ 
for this particular snapshot are very similar to the averages
shown in Fig.1-a and 2-a, clearly indicating incommensurate
behavior in spin and charge. 
As the temperature is raised to T=0.05,
stripes are no longer observed (Fig.3-c) but there are still hole-poor magnetic domains
which produce the (small) incommensurate peak shown in Fig.1-b at 
this density. An
equally weak feature appears in $N({\bf q})$ but it is more difficult to distinguish because
the background raises with increasing momentum,
rather than being flat (Fig.2-b). This may be the reason why
the spin incommensurability is easier to detect than charge
inhomogeneities in models such as Hubbard or t-J, where low temperatures are
difficult to reach. As it can be observed in the snapshot of Fig.3-c
even at T=0.05 hole-rich patches and antiferromagnetically spin aligned
hole-poor domains coexist, suggesting that incommensurate charge and spin order
occur simultaneously due to their mutual interactions in the SF model.
\begin{figure}[thbp]
\centerline{\psfig{figure=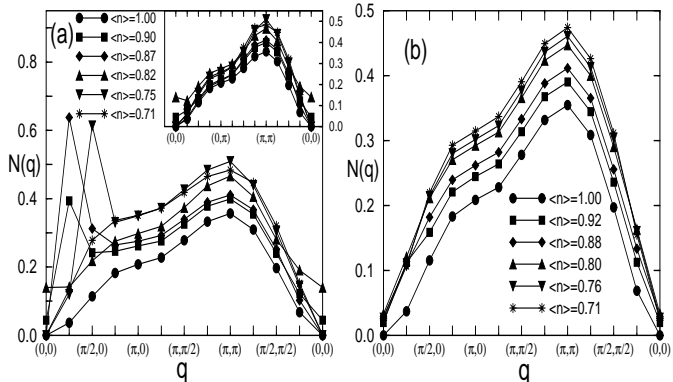,width=9cm}}
\caption{(a) Structure factor $N({\bf q})$ for the electrons versus
 momentum for J=2, $\rm J'$=0.05 and T=0.01 on an 8$\times$8
lattice at several  densities along the path
$(0,0)$-$(\pi,0)$-$(\pi,\pi)$-$(0,0)$ (inset: same as main figure but along 
$(0,0)$-$(0,\pi)$-$(\pi,\pi)$-$(0,0)$); 
(b) same as (a) but at T=0.05. PBC are used.}
\end{figure}
Our results are in qualitative agreement with the conclusions
of Ref.\cite{white,Doug} where it was argued that stripes can be stabilized 
at realistic values of J/t in the t-J model
without the use of
long-range Coulomb interactions.
Moreover, in the SF model here we showed that charge stripes can appear 
spontaneously without the need of using external staggered magnetic fields or
special BC to pin them.\cite{Doug} 
The SF model provides a clean and easy to study framework for
the analysis of stripe formation in models of correlated electrons.
The origin of the stripes in our study can be understood in part  by
analyzing the behavior of $\rm \langle n \rangle$ vs $\mu$
shown in Fig.4-a, where it is observed that the 
density changes rapidly
between 0.5 and 1. At $\langle{\rm n}\rangle$$\approx$0.5 there is a plateau
indicating that this density is particularly stable. For 0.5$\leq \langle{\rm n}\rangle\leq$1 substantial 
charge fluctuations are to be expected due to the large
value of $d\langle {\rm n} \rangle/d\mu$, involving
regions whith density close to 1 (spin domains) and to 0.5
(hole stripes) which correspond to the two highly stable densities that 
appear in the system.
 In fact, 
calculating the charge density along the stripes we
found that 0.5$\rm \leq \langle {\rm n} \rangle_{stripe} \leq$0.65, which seems to
indicate the existence of approximately one hole every two Cu ions.
The charge density on the hole poor regions, on the other hand, has a
value very close to 0.9. These densities inside and outside the stripe
are in excellent agreement with experiments\cite{Tran}.
The SF model improves on
early Hartree-Fock calculations
that predicted a stripe density close to zero.
\begin{figure}[thbp]
\centerline{\psfig{figure=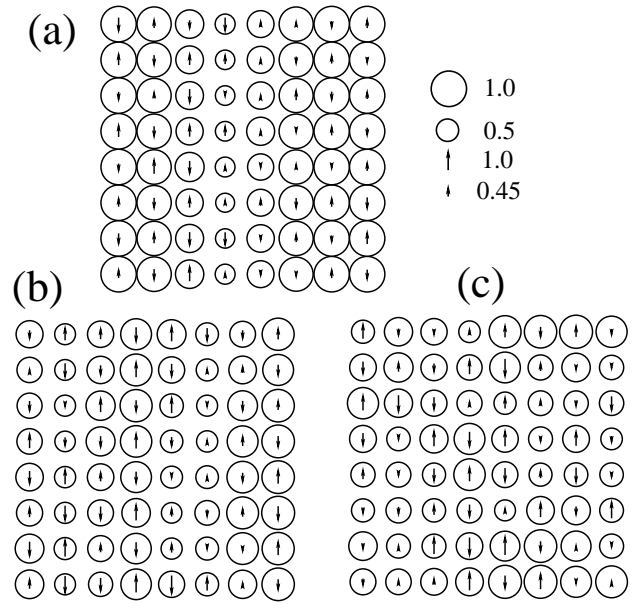,height=8.5cm}}
\caption{ (a) Spin and
charge distribution for a MC snapshot on an 8$\times$8 cluster 
for J=2, $\rm J'$=0.05, T=0.01, and at $\langle {\rm n} \rangle$$\approx$
0.85;(b)  spin and
charge distribution for a snapshot at $\rm \langle {\rm n} \rangle$=0.75 and the
same parameters as in (a). The notation is as in Fig.1-c; (c) same as
(b) but for T=0.05. PBC are used.}
\end{figure}
It is important to note that the SF model
does not phase separate in spite of its large compressibility. 
Using different lattice sizes and BC
we have verified that there is no discontinuity in $\rm \langle {\rm n} \rangle$
vs. $\mu$, while 
the energy presents a
nearly straight line behavior in the density  range
between 0.5 and 1.0 (Fig.4-b). Using 
starting MC configurations  in which all the holes were
together at the center of the cluster it was observed that
 this arrangement decays into
stripes, and comparing the energy of the phase separated and stripe
configurations the
latter was found to have a lower energy than the former.

As remarked before, stripe configurations were observed with several
BC including periodic.
One subtlety encountered in the latter is that for densities where a
single stripe is stabilized, the PBC prevented 
the occurrence of a $\pi$-shift in
the spin domains because it would induce spin frustration 
(Figs.3-a). However, if PBC are replaced by open BC (OBC) 
the stripe still appears and in this case a
 $\pi$-shift is observed (Fig.4-c). 
When the number of
stripes is even, as in Fig.3-b, the $\pi$-shift is spontaneously obtained
independently of the BC. This result is also in excellent
agreement with experiments.\cite{Tran1}

Our simulations can produce dynamical information directly in real-frequency
without the need of carrying out (uncontrolled) analytic extrapolations from
the imaginary axis. This is particularly important to compare theoretical
predictions with
the results of recent
photoemission experiments on optimally doped LSCO which showed
the development of a pseudogap at $\mu$
as the temperature decreases.\cite{endoh} In Fig.4-d we present
the density of states (DOS) for $\rm \langle {\rm n} \rangle$=0.75 at T=0.05,
0.02 and 0.01. A pseudogap at $\omega$=$\mu$ clearly develops for decreasing
temperatures. This is a consequence of the strong density
fluctuations discussed above, and it is similar to the phenomenon recently
observed in the context of manganites where a pseudogap develops due to
the coexistence of hole-rich and hole-poor domains.\cite{gap}
\begin{figure}[thbp]
\centerline{\psfig{figure=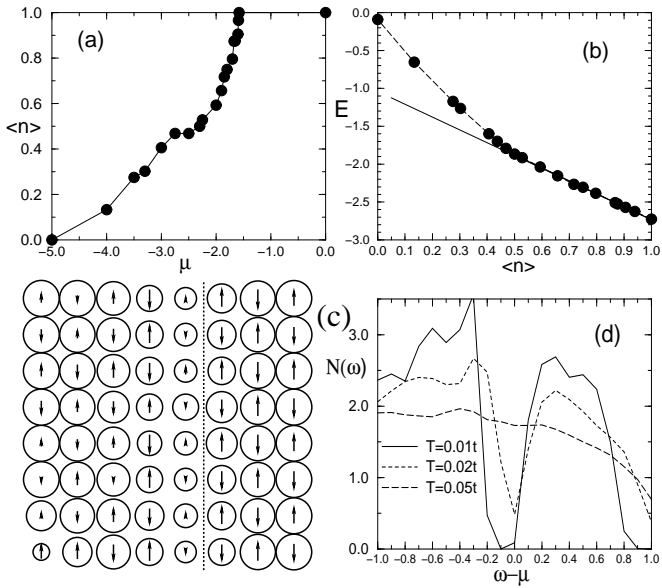,height=8cm}}
\caption{(a) Density $\rm \langle {\rm n} \rangle$ vs $\mu$ for
J=2, $\rm J'$=0.05 and T=0.01 on an 8$\times$8
lattice with PBC; (b) energy versus density for the same parameters as
in (a); (c) spin and
charge distribution for a MC snapshot at $\langle {\rm n} \rangle \approx 0.85$ and the
same parameters as in (a) using OBC. 
The notation is as in Fig.3; 
(d) density of states as a function of $\omega-\mu$ at
$\langle {\rm n} \rangle$=0.75 for different temperatures. The remaining
parameters are as in (a).}
\end{figure}
Summarizing, the SF model has been studied using MC techniques without
$a$-$priori$ assumptions on their properties.
Magnetic and charge incommensurability has been observed upon hole
doping in this model.
The incommensurability is due to the formation of AF domains separated
by metallic stripes of holes. A $\pi$-shift is observed between the domains, and
the stripes are partially filled with an electronic density of about
0.5. Charge and spin incommensurability appear correlated,
and stripe-like configurations are obtained independently of the
BC used and without the long-range Coulomb repulsion. 
The effect arises from the  strong charge
fluctuations between densities 0.5 and 1.0,
which in addition produces a clear
pseudogap in the DOS. Phase separation has not
been observed in this study.
The complex behavior of the SF model reported here, 
not anticipated in previous analysis,
suggests that this model can be as useful for theoretical studies of the
cuprates as  the Hubbard and t-J models, while computationally it
is considerably simpler.

A.M. is supported by NSF under grant DMR-9814350.
Additional support is provided by the National High Magnetic Field Lab 
and MARTECH.


\begin{references}

\bibitem{Cheong} S-W.~Cheong {\it et al.}, Phys.~Rev.~Lett.~{\bf 67}, 1791
(1991); P.~Dai {\it et al.}, Phys.~Rev.~Lett.~{\bf 80},
1738 (1998); H.A.~Mook {\it et al.}, Nature {\bf 395}, 580 (1998).

\bibitem{Tran} J.M.~Tranquada {\it et al.}, Phys.~Rev.~Lett.~{\bf 78}, 338 (1997).

\bibitem{zaanen} D. Poilblanc and T. M. Rice, Phys. Rev. B{\bf 39}, 9749
(1989); J. Zaanen and O. Gunnarsson, Phys. Rev. B{\bf 40}, 7391
(1989).

\bibitem{Eme} V.J.~Emery and S.A.~Kivelson, Physica~{\bf C209}, 597
(1993); ibid {\bf C235} 189 (1994); and references therein.

\bibitem{white} S. R. White and D. J. Scalapino, Phys. Rev. Lett. {\bf
80}, 1272 (1998).

\bibitem{Pines} P.~Monthoux and D.~Pines, Phys.~Rev.~B{\bf 47}, 6069
(1993); A.~Chubukov, Phys.~Rev.~B{\bf 52}, R3840 (1995).

\bibitem{Schrieffer} J.R.~Schrieffer, 
J. of Low Temp.~Phys.~{\bf 99}, 397 (1995);
B.L.~Altshuler {\it et al.}
Phys.~Rev.~B{\bf 52}, 5563
(1995).

\bibitem{Fedro} C.-X.~Chen {\it et al.}
Phys.~Rev.~B{\bf 43}, 3771 (1991).

\bibitem{foot} This is an approximation commonly used in the study of
the SF model. See, for example, Ref.\cite{Pines}. The agreement with
experiments found in
our study provides an extra ($a$-$posteriori$) justification  for this approximation.

\bibitem{adri}  A.~Moreo {\it et al.}, Phys.~Rev.~B{\bf 41} 2313 (1990);
D.~Duffy and A.~Moreo, Phys.~Rev. B{\bf 52} 15607 (1995).

\bibitem{yuno} S.~Yunoki {\it et al.}, Phys.~Rev.~Lett.~{\bf 80}, 845 (1998): 
E.~Dagotto {\it et al.}, Phys. Rev. B{\bf 58}, 6414 (1998).

\bibitem{foot1} The initial conditions when not otherwise specified are
random disordered spin configurations.

\bibitem{yamada}K.~Yamada {\it et al.}, Phys.~Rev.~B{\bf 57}, 6165 (1998).

\bibitem{chino} Y.C. Chen {\it et al.}, Phys.~Rev.~B{\bf 50}, 655 (1994).

\bibitem{Tran1} J.M.~Tranquada {\it et al.}, Nature.~{\bf 375}, 561
(1995).

\bibitem{Zachar} O.~Zachar {\it et al.}, Phys.~Rev.~B{\bf
57}, 1422 (1998). 

\bibitem{Doug} R. Eder {\it et al.}, Phys. Rev. B{\bf 60}, R3716 (1999);
S.~White and D.~Scalapino, preprint, cond-mat/9812187.

\bibitem{endoh} T.~Sato {\it et al.}, preprint, to appear in Phys. Rev. Lett.

\bibitem{gap} A.~Moreo {\it et al.}, to appear in Phys.
Rev. Lett.

\end{references}
\end{document}